# Reflection-in-Action Markers for Reflection-on-Action in Computer-Supported Collaborative Learning Settings

Élise Lavoué [1,2]

[1] Magellan, IAE Lyon, Université Jean Moulin Lyon 3, France

[2] LIRIS, UMR5205

Postal address: IAE Lyon, 6 cours Albert Thomas BP 8242, 69355 Lyon Cedex 08, France

E-mail: elise.lavoue@univ-lyon3.fr

Gaëlle Molinari [3]

[3] Distance Learning University Switzerland & TECFA, University of Geneva

Postal address: Techno-Pôle 5, Case Postale 218, CH-3960 Sierre, Switzerland

E-mail: gaelle.molinari@unidistance.ch

Yannick Prié[4]

[4] University of Nantes, LINA - UMR 6241 CNRS, France

Postal address: Rue Christian Pauc, BP50609, 44306 Nantes cedex 3, France

E-mail: yannick.prie@univ-nantes.fr

Safè Khezami[5]

[5] Université de Toulon, I3M

Postal address: Campus Porte d'Italie, 70 Avenue Devoucoux, 83000 Toulon, France

E-mail: safa-khezami@etud.univ-tln.fr

**Abstract**: We describe an exploratory study on the use of markers set during a synchronous collaborative interaction (reflection-in-action) for later construction of reflection reports upon the collaboration that occurred (reflection-on-action). During two sessions, pairs of students used the Visu videoconferencing tool for synchronous interaction and marker setting (positive, negative or free) and then individual report building on the interaction (using markers or not). A quantitative descriptive analysis was conducted on the markers put in action, on their use to reflect on action and on the reflection categories of the sentences in these reports. Results show that the students (1) used the markers equally as a note-taking and reflection means during the interaction, (2) used mainly positive markers both to reflect in and on action; (3) paid more attention in identifying what worked in their interaction (conservative direction) rather than in planning on how to improve their group work (progressive direction); (4) used mainly their own markers to reflect on action, with an increase in the use of their partners' markers in the second reflection reports; (5) reflected mainly on their partner in the first reflection reports and more on themselves in the second reports to justify themselves and to express their satisfaction.

1. Introduction

Computer-Supported Collaborative Learning (CSCL) research has shown the need to scaffold collaboration so as to ensure that learners benefit from working together (Dillenbourg, 1999). In the CSCL field, the awareness approach is used to support collaborative learning by monitoring and regulating the interaction between learners. This approach is technology-based, and consists of providing information about group members' knowledge, emotions, actions and interactions during collaborative learning (e.g. Molinari, Chanel, Bétrancourt, Pun, & Bozelle, 2013; Sangin, Molinari, Nüssli, & Dillenbourg, 2011). Such awareness information is supposed to help learners reflect on how they work together and understand how to improve their group performance. Within the awareness approach, there are still few tools designed to encourage learners' reflection during ("reflection-in-action") as well as after their interaction with their partners ("reflection-on-action") (Schön, 1987).

In this paper, an exploratory study is reported in which students in Psychology used the Visu tool – which is a tool for both reflection-in and on-action – in CSCL settings. Visu is a web videoconferencing platform (Bétrancourt, Guichon, & Prié, 2011) that allows participants to take notes and report their feelings about their interaction with their partner at any time during remote synchronous collaboration. More precisely, in this study, students could take notes using two types of markers: emotional markers to express either negative or positive feelings about the way they collaborate; and non-emotional markers to provide any other types of comments on the on-going activity. Visu also provides students with the possibility to later review the traces of their group's work (audio/video recordings of the interactions, as well as self and partner's markers with their associated notes). Such review can lead to the production of self-reflection reports, as in this study.

This paper focuses on students' reflection processes regarding their remote collaborative work. More particularly, we are interested in the way they used the markers and the associated notes their partner or themselves did create during synchronous interaction sessions, to individually reflect upon the quality of their collaboration after it had taken place. In this study, students were organized in dyads, and each dyad was involved in two consecutive synchronous CSCL sessions. After each of these sessions, students were asked to individually produce a reflection report. They were instructed to organize their reports into two parts, a retrospective part in which they had to describe their perceptions regarding the quality of the interaction they just had with their partner, and a prospective part in which they had to think about how to improve their work as a team.

The general research objective of this study is to describe the use of one's own- and partner's emotional and non-emotional markers – that are set during interaction to take notes and reflect on the collaborative learning process (reflection-in-action) – in later reflection-on-action. More precisely, our questions are: what kinds of markers (emotional or non-emotional) did students use to reflect-in-action while interacting with their partner? What kinds of markers (emotional or non-emotional, own or partner's) did students use to reflect-on-action when building their self-reflection report after collaboration? What kinds of reflection-on-action processes (retrospective processes, e.g. evaluation, causal attribution, affective reactions) and prospective processes (e.g. task analysis, motivational beliefs) – see (Zimmerman, 2002) – were related to the use of emotional and non-emotional markers as well as to the use of own- and partner's markers? To what extent did such reflection processes vary from the first synchronous CSCL session to the second with the use of the different types of markers?

In Section 2 we define the theoretical background of this research as well as the existing tools and platforms used to support reflection processes in CSCL settings. Section 3 deals with the Visu platform, which provides collaborative learners with the possibility to self-report what they are experiencing during interaction, including cognitive and affective information about themselves, their partner and the group as a whole. In Section 4 we present the study we conducted in an ecological context, namely during the Educational Psychology Course of the Bachelor of Science in Psychology at the Distance Learning University Switzerland. We finally sum up the main results of this study and present our future works.

2. Related Background

*2.1. Theoretical Framework for Analyzing Reflection in CSCL*

*Regulation in CSCL.* Although there is a growing body of research that focuses on socially shared metacognition and regulation e.g. (Iiskala, Vauras, Lehtinen, & Salonen ,2011; Järvelä & Järvenoja 2011), we still know little on how learners regulate and reflect upon their own activity, their partner's activity as well as their group activity in CSCL settings. Regulation is defined as controlled processes through which thoughts, emotions, strategies, and behaviors "are oriented to attaining goals" (Zimmerman, 2002). Regulation processes have proved to be important for successful collaborative learning (Järvelä & Järvenoja, 2011). They occur mainly in episodes when collaborating partners are confronted with conceptual or relational difficulties, and their role is either to facilitate or to inhibit representations and activities (Iiskala et al., 2011). Individuals can engage in three types of regulation processes during collaborative learning tasks. First, they can individually reflect upon how to regulate their learning processes and outcomes (self-regulation). Second, they can reflect upon how to help and support their partners in their learning (other-regulation). As pointed out by Järvenoja (2010), although other-regulation is beneficial to the whole group, it can be viewed as a form of individual regulation as it may be used (at first) for personal goals. Third, regulation and reflection can also be co-constructed processes (shared regulation): learners can discuss and develop together common strategies to control their group activity and the learning challenges they face during interaction.

According to Van der Puil, Andriessen, and Kanselaar (2004), regulation processes in collaborative learning situations can take two directions: conservative or progressive. In the conservative direction, regulation can be seen as a "looking-backward" process through which group members reflect on what was right or wrong with their working relationship (social regulation) as well as with the way they shared and negotiated knowledge (cognitive regulation). In the progressive direction, regulation is viewed as a "looking-forward" process through which collaborators pay attention on how to achieve the learning task goals in the future. Van der Puil et al. (2004) also showed that the way group members regulate their work could be dominated by conservative forces; in such cases, they would be mainly focused on repairing the relation, relegating to a second plane the learning and task goals.

*Reflection as a regulation process*. Reflection is considered as one phase of regulation in the models proposed by Pintrich (2004) and Zimmerman (2002) to describe self-regulated learning (SRL). More precisely, reflection refers to cognitive and affective processes that take place once the overall task or part of the task has been completed. In this phase, learners assess the quality of work being performed (evaluation), and try to explain successes and failures (causal attribution). They also positively or negatively react to such attributions (affective reactions). They can affectively react to the collaborative situation, by expressing different levels of satisfaction (satisfaction/affect), protecting the feeling of competence or proposing adjustments and changes in behavior necessary to succeed (adaptive/defensive decisions). In CSCL settings these reflection processes (evaluation, causal attribution, satisfaction/affect, adaptive/defensive decisions) can be individual or collaborative, and may focus on oneself, the partner, the group, the task or the context. Reflection is considered as crucial for learning as it helps individuals to internalize and reconstruct what they have (socially) learned, and to transfer their knowledge and skills.

Both SRL models (Pintrich, 2004; Zimmerman, 2002) also identified two other phases of regulation, namely the forethought and performance phases. The forethought phase refers to processes that are carried out in preparation for the task. In this phase, learners analyze the task (task analysis), establish goals (goal setting) and plan strategies to achieve them (strategic planning). They also activate motivational processes (motivational beliefs) such as efficacy and task interest/value beliefs. The performance phase refers to processes that occur during the task. In this phase, learners actively keep track of the progress of the task, and activate strategies to maintain their concentration and motivation. The monitoring and control processes in Pintrich's (2004) model are included in the performance phase described in Zimmerman's (2002) model. As for reflection processes, one may expect that processes involved in the forethought and performance phases (i.e., task analysis, motivational beliefs, monitoring, control) can be self-regulated, other-regulated or socially shared-regulated learning processes. Finally, SRL models assume that in each phase (forethought, monitoring, control, reflection and reaction) the activities of regulation concern four aspects of learning: cognition, motivation/affect, behavior and context (Pintrich 2004).

In both Pintrich's (2004) and Zimmerman's (2002) models – see also (Kolb, 1984) – SRL processes occur in a cycle in which individuals first act, then reflect back upon their experiences, assimilate their reflections in a theory and deduce implications for future actions from that theory. In other words, in these models, self-regulation phases are time-ordered: reflection happens after the performance, and before planning and goal setting. Schön (1987) assumed, however, that reflection can occur both during (and in) the task being performed (reflection-in-action) and after the task, e.g. when mentally replaying it (reflection-on-action). According to Schön (1987), reflection-in-action is a process activated when something different, unusual or even inappropriate suddenly happens and claims attention. Real-time adjustment and modification of actions can then arise from reflection-in-action. Boud, Keogh, and Walker (1985) described reflection-on-action as consisting of three elements: (a) going back to a past experience; (b) re-evaluating it in the light of current insights and knowledge, and with particular focus on its emotional aspects; and (c) deriving new perspectives for future activities from this evaluation. These elements are quite similar to those included in the self-reflection loop described in SRL models through which learners regulate their learning behavior based on cognitive judgments, affective reactions and task/context evaluations (Pintrich, 2004). The outcome of reflection-on-action can be therefore cognitive, affective and/or behavioral, including the planning and implementation of changes.

*Emotional and non-emotional markers for reflection*. In the present study, we analyzed how learners used markers both during two synchronous interaction sessions and after each session to build an individual reflection report about their collaborative work. During synchronous sessions, markers could be used to take notes and underline relevant information shared between learners during interaction (attention/note-taking markers), but also to reflect upon the on-going work (reflection-in-action markers). Two types of markers were distinguished in this study, namely non-emotional markers (free markers) and emotional markers. Emotions experienced in individual and collaborative learning settings can be either positive or negative, and focus either on the learning activity or on learning outcomes (Pekrun, 2006). According to cognitive appraisal theories, emotions in learning are described as the result of evaluation based on different criteria such as the perceived level of control over the task, or the perceived value of the situation (Pekrun, 2006; Scherer, 2005). Emotions are recognized as having considerable impact on cognitive, motivational and regulatory processes involved in learning (D'Mello, & Graesser, 2012; Kort, Reilly, & Picard, 2001; Pekrun, 2006). Moreover, collaborative learning experience is characterized by continuous fluctuations of emotions within and also between learners. In CSCL settings, learners benefit from being aware of what their collaborative partners feel during interaction. Learners that communicate their emotions to each other are more likely to build on their partner's ideas and to interact together in a transactive way (Molinari et al., 2013). In the present study, we decided to give learners the possibility to use positive and negative markers, since those emotional markers could facilitate learners in their evaluation processes and in identifying successes and failures during their interaction.

After each interaction session of the present study, the markers and their associated notes could be used as "anchor points" for learners' individual reflections on the quality of their collaboration (reflection-on-action markers). We used Zimmerman's (2002) model as a framework to analyze the content of self-reflection reports. More precisely, sentences in the reports were analyzed as referring to either the reflection phase (evaluation, causal attribution, satisfaction/affect, adaptive/defensive decisions) or the forethought phase (goal setting, strategic planning, efficacy beliefs, outcome expectations, task value, interest and goal orientation).

*2.2. Computer Tools for Reflection in CSCL*

Students can reflect on individual or collective experiences, "in isolation or in association with others" (Boud et al., 1985, p. 19). Different types of computer tools can support reflection processes: feedback tools, group awareness tools, and regulation tools. In this section, we study these tools according to three characteristics of the Visu platform we used for our study (see section 3): the subject on which the reflection is focused (self, others and/or the group); the time of the reflection (synchronous, i.e. reflection-in-action, or delayed, i.e. reflection-on-action); and the type of reflection (cognition, motivation/affect, behavior, and context, cf. (Pintrich, 2004)).

*Feedback tools for reflection.* According to Kluger and DeNisi (1996), feedback is information provided to increase performance. Self and peer assessment are a form of feedback often used for formative assessment, and have been found to foster students' reflection on their own learning process and learning activities (Dochy, Segers, & Sluijsmans, 1999). In most of current computer-supported learning environments, the information about students' performance is automatically calculated and given back to them immediately after their learning activity. Biesinger and Crippen (2010) gave an automatic feedback based on students' quiz scores so as to support learners' goal orientation, self-regulation, self-efficacy and achievement processes. They proposed two types of feedback to individual learners: comparison with their own prior attempts (quiz average), and comparison with the group (class quiz average). Zou and Zhang (2013) aimed at promoting students' self-regulated learning, by presenting them with the outcome of their activity (overall scores to tests, sub-scores to each topic, percentile position). Students were also provided with a feedback on their self-performance and learning process so as to be able to evaluate themselves their self-regulation strategy use. Feedback tools have also been used with other peer scaffolding tools, for instance peers' votes, annotations and notes as explicit scaffolding messages, as in the KnowCat platform for stimulating collaborative learning (Pifarre & Cobos, 2010).

All these feedback tools have proven to be useful to enhance individuals' reflection during the learning activity (reflection-on-action), mainly by comparison with others' performance. These tools are not designed to provide information on the collaborative processes of the group. Furthermore, the feedback focuses only on cognition, and do not provide information on the motivation/affect, behaviors, and context.

*Group awareness tools for reflection*. In CSCL settings, it can be rather difficult for learners to construct a clear and precise understanding of what their partner feels, does or intends to do when relevant cues are missing (e.g. non-verbal: gestures, eye gaze, etc.; or social context-related: geographic, organizational or situational information). The absence of such cues can impair learners' awareness about their own- and their partner's activities (Dourish & Bellotti, 1992), social interaction and communication effectiveness (Kreijns,

Kirschner, & Jochems, 2003). Group awareness technologies (Buder, 2011) are designed to circumvent the lack of awareness information in computer-mediated collaboration. Such technologies aim at collecting data about users' characteristics and behaviors during collaboration, and reflecting this information back to them. Group awareness covers the perception of behavioral, cognitive, and social context information on a group or its members (Bodemer & Dehler, 2011), and group awareness tools generally focus on one of these types of information. Kimmerle and Cress (2008) proposed a tool that provides behavioral information to the learner, such as the number of his/her contributions in comparison with that of other members or the whole group. Their study showed that information concerning the contribution behavior of individuals clearly increased their cooperation rate in comparison to those receiving no feedback and those merely receiving group feedback. Janssen, Erkens and Kirschner (2011) also showed that social information, such as participation levels during online discussions, can stimulate learners to participate more in online discussions and collaborative processes. Lajoie and Lu (2012) provided learners with a structured template for collaboratively constructing, annotating and sharing documents, which enhanced metacognitive activity and led to effective forms of co-regulation (planning and orienting). Cognitive information was also presented with a positive impact on learning outcomes, as in the Knowledge Awareness Tool (KAT) (Sangin et al., 2011) where members were shown a virtual representation of their peer's level of prior knowledge. In project-based learning, cognitive information focusing on the tasks to carry out according to the project and learning goals enhanced self-regulated learning processes (self-monitoring and self-judgment) (Michel, Lavoué, & Pietrac, 2012).

  As a conclusion, group awareness tools implicitly guide learners' behavior, communication, and reflection by presenting information on the cognitive and/or social behavior of the others (the learning partners) or the group. But, according to Prins,

Sluijsmans, and Kirschner (2006), providing group members with this information is not enough to positively alter their behavior. Group members also need to process this information and ask themselves whether they understand, accept, and agree with the feedback. In other words, they must reflect upon the feedback (Phielix, Prins, Kirschner, Erkens, & Jaspers, 2011). Group awareness tools do not provide learners with appropriate means to support this reflection during (reflection-in-action) and/or after (reflection-on-action) the collaborative activity and so be really aware of the metacognitive skills they applied.

*Regulated learning platforms in CSCL*. In recent years, several platforms (set of tools) have been developed to provide learners not only with feedback information, but also with the appropriate means to reflect on this information, so as to help them adjust their goals and their strategy to attain them. These platforms can be named Self-Regulated Learning (SRL) platforms, as they support learners in all the steps of self-regulated learning processes. For that, they provide tools that support both reflection-in-action, by displaying information during the activity on the present situation, and reflection-on-action to encourage users to engage in delayed reflection after their activity. For instance, Study Desk (Narciss, Proske, & Koerndle, 2007) is composed of various learning resources, monitoring tools and tutoring feedback to initiate task and content-related learning activities (marking, note-taking and elaboration) and meta-cognitive activities (monitoring and evaluating the learning process and outcomes). The MetaTutor platform (Azevedo, Witherspoon, Chauncey, Burkett, & Fike, 2009; Azevedo et al., 2012) offers an adaptive scaffolding and feedback provided by a human tutor or a pedagogical agent that leads to greater deployment of sophisticated planning processes, meta-cognitive monitoring processes, and regulation during learning. More specifically, some tools were developed to lead learners to express their reflection, for instance on the form of a reflective journal (Yang, 2010), or to explain their learning processes. For instance, Betty's Brain (Roscoe, Segedy, Sulcer, Jeong, & Biswas, 2013) encourages students to produce explanations of their emerging understanding via a causal concept maps that allows an agent system to give them prompts. Aleven and Koedinger (2002) designed a cognitive tutor to enhance students' self-explanations skills by providing them with assistance in the form of hints on how to self-explain as well as feedback on their explanations. The platforms and tools we presented are used to support self-regulated learning processes for individuals involved in a learning activity, by providing them means for self-reflection in and/or on action.

A few self-regulated learning platforms were specifically developed to gain a set of critical skills needed to engage in and self-regulate collaborative learning experiences by supporting reflection on the others and on the group. For instance, Metafora (Dragon et al., 2013) is composed of planning, reflection and discussion tools, to help groups of learners develop reflection on the group learning processes. gStudy (Hadwin, Oshige, Gress, & Winne, 2010) provides students with chat tools, objects sharing, note templates and coaching to support three types of regulation: self-regulation, co-regulation and shared-regulation.

However, current self-regulated learning platforms oriented towards collaborative contexts mostly lack social and emotional feedback. Indeed, most CSCL environments focus on supporting cognitive or task-related processes and limit possibilities for social or non-task-related processes (Kreijns et al., 2003). According to Phielix et al. (2011), the absence of visual, non-verbal cues can cause specific communication and interaction problems since there are few possibilities to exchange socio-emotional and affective information. Phielix et al. (2011) propose a platform that combines a tool for reflection-in-action (Radar) and a tool for reflection-on-action (Reflector). The Radar tool presents learners with anonymous information on six traits on their cognitive (productivity and quality of contribution) and social (influence, friendliness, cooperation, reliability) behaviors. The Reflector tool is designed to ask group members to reflect upon their individual behavior and their group's past and future performance during a collaborative writing task. A positive effect of the use of these two tools on the level of group process satisfaction has been found. Reflector also encouraged participants to consider improving group performance as an explicit goal. To our knowledge, the platform developed by Phielix et al. (2011) is the only one focusing on the perceived social and cognitive behavior of the group during the two times of the reflection. However, the information presented to learners on the Radar tool is only related to themselves and not to the collaborative processes.

As a conclusion, we observe that most of the feedback tools and self-regulated learning platforms have been designed to enhance learners' reflection on their individual experiences with a focus on themselves. Only group awareness tools and a few SRL platforms can foster reflection on the collaborative processes with a focus on the others and the group. Feedback tools and group awareness tools do not provide support for learners to explain their performance and the learning processes and so to explicitly reflect in and/or on action. Some self-regulated learning tools and platforms have been specifically developed to support these two times of the reflection. A few of them support reflection on collaborative processes of the group, but they focus on the cognitive aspect and do not give information on group members' affect and motivation. To our knowledge, the platform developed by Phielix et al. (2011) is the only one to focus on this aspect, but the feedback given to students is about their own individual behavior and not on the collaborative processes (others and/or the group). In next section, we present the Visu platform, which support self-regulated learning processes in a collaborative context, by providing learners with information on themselves, their partner and the group as a whole, including information on their affect and motivation.

3. Visu: a tool for Synchronous and Delayed Reflection on collaborative interaction

The Visu tool used in the present study is a video conferencing tool that allows both reflection-in-action (by setting up markers and notes during the interaction) and reflection-on-action (by allowing retrospective analysis of the interaction using markers and notes, as well as report building). The availability of positive and negative markers explicitly permits to reflect on emotional aspects of the experience. The capacity of the tool to record the whole interaction, but also to share markers fosters reflection not only on the individual, but also on the partner and the group as a whole. Is this section we first introduce the original design rationale of Visu, before describing the interface of Visu 2 and the improvements we made for this study.

Visu was designed and built inside the ITHACA project[1], with the objective of taking further existing practices for teaching live distant tutoring to FLE (French as a foreign language) apprentice tutors by leveraging the use of markers and recordings in video-based synchronous collaborative systems (Clauzel, Sehaba, & Prié, 2010). Existing practices aimed at improving online tutors' competence through reflective analysis (Guichon, 2009). More specifically, they were based on tutoring sessions that made use of the Skype video-conferencing tool between Lyon and Berkeley universities. One student dyad had to prepare the activities and the associated material for each week's French language tutoring session. This dyad videoconference with another dyad of foreign students was then filmed with an external camera, and the video was distributed to them on a DVD that they had watch and comment as non-guided self-confrontation (Guichon, 2009). A debriefing session was then held with the whole class to discuss how the interaction unfolded and reflect on their practices. Based upon that experience, the need was identified to build a system that would facilitate these practices of reflection-on-action, but also permit new ones, such as individual and group reflection-in-action, and easier sharing of reflections. Visu was then created as a videoconferencing tool with specific features that allowed to prepare interaction outlines, to take notes on the interaction during its unfolding while recording the whole interaction (video, actions on the interface, markers and associated notes), and to reuse this recording later for retrospective activity (Bétrancourt et al., 2011).

---

[1] Interactive Traces for Human Awareness and Collaborative Annotation (2008-2011).

Visu 2 evolution added numerous ergonomic enhancements, as well as several features such as markers and notes sharing, sharable reports building, etc. Visu 2 was used both for language teaching tutoring (as Visu 1, with an asymmetric relation between tutors and trainees), but also –as in the experiment we describe in this article– for more collaborative activity. For this, we tailored the tool on aspects regarding 1/ the predefinition of two markers (red and green) to account for the affective/judgment dimension, 2/ the sharing of markers and notes between participants after the interaction to provide support for reflection on the group activity, and 3/ the possibility to build a report using one's and other participant's markers and notes. For the sake of clarity, let us notice that there were six possibilities to reflect-in-action during the interaction: set up a positive marker (green); set up a negative marker (red); set up a positive (respectively a negative) marker with a textual note; set up a free marker with a note; set up a free marker without a note (of course, this latter case would just indicate a moment in the interaction, and be mostly useless without further explanation).

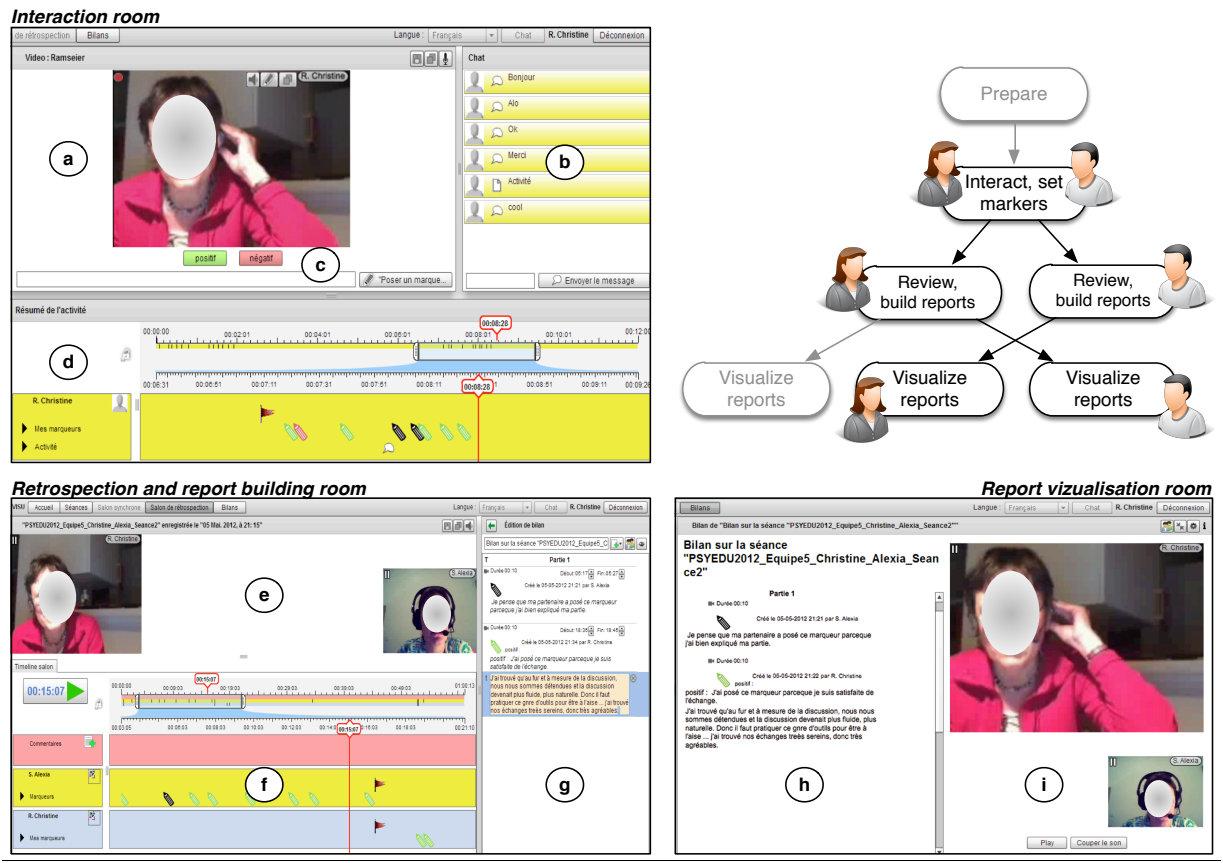

Figure 1. The three Main Rooms of Visu 2 and its General Workflow

We can now present with more details the Visu 2 tool that was used during the experiment described in this paper. Let us begin with the general workflow of Visu on the upper right Figure 1. First, a user (e.g. a tutor) can *prepare* the interaction material (this feature was not used here). Second, the participants *interact* synchronously; they can *set markers* and *notes* during the collaborative activity. Third, after the interaction, they can access the recordings of the interaction, *review* them and prepare rich media *reports* to share. Last, users who have access to reports can view them (here the participants only accessed to their partner's report). Figure 1 also presents the three main rooms of Visu that are of interest here[2].

The *interaction room* (upper left) is organized around classical videoconference features, namely a video zone (a) and a chat (b). When the video recording is on, users can leave markers on a horizontal timeline (d). To do so, they can either put a free marker by defining a note and then hitting return (or the button "Poser un marqueur") or use buttons linked to predefined emotional markers: positive (green button) or negative (red button) (c). Such emotional markers are not associated to a note; should users wish to add a note, they can do so by using the textual form before clicking on the green or red button. The markers set by users are not visible by their partners during the course of collaboration (as it was thought this would affect the quality of interaction by focusing too much users' attention on their partners' markers). In Figure 1, free markers appear in black color while positive (respectively negative) markers are in green (respectively red). Apart from markers, the timeline (d) can also contain representations of the various actions of the user and its partners, serving both as an history and an awareness backchannel for tutors, e.g. for trainees' document consumption.

---
[2] We skip the presentation of the *Preparation room*.

The *retrospection and report building room* (lower left) can be accessed at any time after the interaction so as to review the recordings of the synchronous session (videos, markers and notes). The markers and associated notes left during the interaction appear on the horizontal timeline (f) and now all the markers and notes are visible for users who participated in the interaction. Reviewing the interaction mainly means watching and listening to the videos (e), mostly by using the timeline for navigation. Users can then individually build a reflection report on the collaboration process, by using any of the markers and notes that were set during the interaction. For this, they can either drag and drop markers in the editing space of the retrospection room and modify the text of the notes as they wish or create text blocks from scratch (g).

The *report visualization room* (lower right) present reports composed of several blocks (h). These blocks can be titles; simple texts; interaction video fragments (possibly with textual comment, possibly presented with the colored markers that may have been used to create them); or new audio recordings (possibly with textual comment). In the experiment, users were asked to use only titles and simple texts.

Reflection is present at every level of these rooms: interaction room provides in-action reflection based on its timeline, retrospection room is designed for reflection, both by navigating the recordings and by building reports; report visualization room allows to make use of second hand reflection material. In our study, we focused on the reflection processes that occurred in the interaction and retrospection room, as the use of VISU corresponds to the three phases of the Zimmerman's model we adopted: use of reflective markers and associated notes in the online interaction room (performance phase), and use of interaction traces (videos and markers) in the retrospection room (self-reflection and forethought phases). In comparison with existing reflection tools, VISU provides learners with information on their affect and motivation (on the form of markers and associated text). As all markers (its own and those of the partner) put during the synchronous collaborative sessions are visible in the retrospection room, this information can be about the learner, the partner and the group as a whole.

4. Study

*4.1. Context, Participants and Procedure*

The study took place in an ecological context, namely during the educational psychology course of the Bachelor of Science in Psychology at the Distance Learning University Switzerland. This course is a semester course divided in (a) 5 three-week online classroom periods and (b) 5 one-day face-to-face classrooms. Each of the online classroom periods is dedicated to one topic in educational psychology (period 1: key concepts in learning and teaching, period 2: behaviorism, period 3: cognitivism, period 4: constructivism/socio-constructivism, and period 5: collaborative learning). Ten students (9 women and 1 man; mean age of 35 years) participated in this educational psychology course in the 2011-2012 academic year. All these students came from very different professional backgrounds.

The study was carried out during the 4th three-week online classroom period (constructivism/ socio-constructivism). In this period, students were asked to work in dyads (5 teams) and used the Visu platform during two synchronous collaborative sessions; the 1st session was held during the 1st week of Period 4, the 2nd session during the 2nd week. During these CSCL sessions, students were invited to discuss and share their understanding about four introductory texts on Piaget's and Vygotsky's theories of learning (two "Piaget" texts and two "Vygotsky" texts). A CSCL script, inspired from a Jigsaw (macro) script developed by Buchs (2002), was used to organize both CSCL sessions. In this script each member of the dyad was invited to read a text in preparation for each sessions: student 1 read the "Piaget" Text 1 for Session 1 and the "Vygotsky" Text 2 for Session 2, while student 2 read the "Vygotsky" Text 1 for Session 1 and the "Piaget" Text 2 for Session 2. Each student then depended on the other to access the content of the two texts (s)he had not read. Both CSCL sessions were composed of three consecutive collaborative phases:

1. Explanation phase 1 (15 minutes): in Session 1, student 1 took the role of teacher and explained the "Piaget" Text 1 to student 2 who took the role of listener/questioner (in Session 2, student 2 explained the "Piaget" Text 2 to student 1);
2. Explanation phase 2 (15 minutes) where students exchanged their previous roles: in session 1, student 2 took the role of teacher and explained the "Vygotsky" Text 1 to student 1 who took the role of listener/questioner (in Session 2, student 1 explained the "Vygotsky" Text 2 to student 2);
3. Comprehension test (30 minutes): both students were provided with two comprehension questions (one question per text) that they had to answer together orally.

The two synchronous sessions took place in the VISU interaction room. Learners were asked to set markers at any time they want during interaction, with or without associated textual notes. As presented in Section 3, three predefined markers were available, two emotional markers – a red/negative marker and a green/positive marker – and one non-emotional marker – a black/free marker. Students received the instructions to use red/negative makers in moments when they perceived the interaction with their partner as unpleasant or when they experienced some difficulties to understand what their partner said. They were also asked to use green/positive markers to indicate pleasant or meaningful moments of interaction with their partner. Students were also instructed that they could not access the markers set by their partner during the interaction, but that both their own- and their partner's markers would be available after their group work for helping them self-reflect on what did work or not regarding the interaction with their partner.

After each CSCL session, students were asked to write an individual report about their teamwork using the editing space of the VISU retrospection room (see Figure 2 for an excerpt of a report). The report had to be composed of two parts, a *retrospective* part and a *prospective* part. For this task, students were provided with the following instructions: "In the retrospective part, we ask you to express your personal perception on how your partner and yourself have collaborated. This part should concern your own activity, your partner's activity as well as the work of your team. In the prospective part, you have to think about how to improve your team's work, in particular, your collaborative strategies and the quality of the relationship with your partner". Students were asked to integrate markers (their own and their partner's) in the report with a drag and drop process as well as write new text blocks. When reusing their own markers, they were instructed to explain when and why they set them during interaction. They also had to comment on the partner's markers they reused.

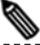

Figure 2. A Reflection Report composed of Markers and Sentences. The original French textual content has been translated on the right.

*4.2. Research Questions and Analysis Method*

On the one hand, we analyzed how learners used emotional (positive, negative) markers and non-emotional (free) markers during their synchronous collaborative learning activity. Learners could use markers and associated notes (1) to make a real-time assessment of their group work (reflection-in-action markers associated with reflection-focused notes), or (2) as "external retrieval" cues for helping them remember relevant information about the knowledge domain or about the management of the tool/task during interaction (note-taking markers associated with content-focused or tool/task-focused notes). On the other hand, we focused on reflection-on-action processes in CSCL settings. We examined how learners used their own- and their partner's emotional and non-emotional markers to individually reflect upon their own work, their partner's work and their group work after collaboration. We also analyzed the sentences of the retrospective and prospective parts of reflection reports using categories based on Zimmerman's (2002) self-regulated learning model.

More precisely, with respect to markers and associated notes produced during the two synchronous CSCL sessions, our research questions were as follows: 1) What types of markers (non-emotional/free, emotional/positive, emotional/negative) did learners set during interaction? 2) To which extent were emotional and non-emotional markers associated with notes? 3) How did the use of emotional and non-emotional markers vary from the first CSCL session to the second? 4) For which purpose (reflection-in-action, note-taking) did learners use emotional and non-emotional markers during interaction?

With regard to the markers and associated notes that were integrated into the reflection reports, and to the sentences of the retrospective and prospective parts of reports, our research questions were as follows: 5) To which extent did the use of the partner's markers differ from the use of personal markers when building self-reflection reports on the collaborative work? 6) What types of personal and partner's markers (non-emotional/free, emotional/positive, emotional/negative) did learners use in their reports? 7) What kinds of reflection processes (reflection and forethought processes) were involved when reviewing the interaction with the partner, and towards whom (the learners themselves, their partner or their group) were such processes directed? 8) What kinds of reflection processes were related to the use of emotional (positive, negative) and non-emotional (free) markers, as well as to the use of personal and partner's markers in learners' reports? 9) How did the use of markers as well as the related reflection processes vary from the first report to the second?

To answer these questions, we performed a quantitative descriptive analysis given that the number of dyads was relatively small ($N = 5$). For the synchronous CSCL sessions ($1^{st}$ and $2^{nd}$ sessions), the analysis was performed on the following measures: 1) The number and type of markers set (red/negative, green/positive and black/free markers), 2) The number and type of markers associated or not with notes, 3) The type of textual notes created and associated to markers (tool/task-focused, content-focused or reflection-focused notes). Two researchers encoded the textual notes ($N = 89$) with an agreement rate of 94,4%. With respect to this high agreement rate, we considered one of the two encoding for the analysis.

For the reflection reports (R1 after the 1$^{st}$ CSCL session and R2 after the 2$^{nd}$ session), we focused on the number of markers used in reports depending on their author (personal/self markers and partner's markers) and their type (red/negative, green/positive and black/free), as well as on the reflection category to which each sentence of reports referred. We distinguished between two main categories. The first one referred to the *reflection* phase and consisted of 4 sub-categories: EV = evaluation, CA = causal attribution, SA = satisfaction/affect, and AD = adaptive/defensive. The other one referred to the *forethought* phase and consisted of 6 sub-categories: GS = goal setting, SP = strategic planning, EF = Efficacy, OE = outcome expectations, IV = intrinsic value, and LO = learning goal orientation. Two researchers encoded the sentences of the reports ($N = 310$) with an agreement rate of 92,6%. With regard to this high agreement rate, we considered one of the two encoding for the analysis.

*4.3. Results*

*4.3.1 Markers and Associated Notes during Synchronous Collaborative Sessions*

We analyzed the number and type of markers (green/positive, red/negative, black/free) created during the two collaborative sessions (Sessions 1 and 2). Moreover, we examined the extent to which markers set during interaction were associated or not with textual notes, and also the change in the use of markers with or without notes between the two sessions. Finally, we analyzed whether notes produced in association with markers were used to either self-reflect (reflection-focused notes), to emphasize relevant learning content (content-focused notes) or to report difficulty using the tool or managing the task (tool/task-focused notes).

*Number and Types of Markers with or without Notes.* In both sessions, the highest percentage was for green/positive markers followed by black/free markers. The lowest percentage was for red/negative markers (see Table 1). Moreover, there was a decrease in the overall number of markers set during interaction between Sessions 1 ($N = 129$) and 2 ($N = 74$). Table 1 shows that this decrease concerned the three types of markers. The percentage of red/negative markers decreased over both sessions while the percentage of green/positive markers increased. The percentage of black/free markers remained constant.

|  | Session 1 | | Session 2 | | Change |
|---|---|---|---|---|---|
|  | No. | % | No. | % | % |
| Red/Negative markers | 21 | 16% | 8 | 11% | -5% |
| Green/Positive markers | 67 | 52% | 42 | 57% | +5% |
| Black/Free markers | 41 | 32% | 24 | 32% | 0% |

Table 1. Number and Percentage of Red/Negative, Green/Positive and Black/Free markers set during Synchronous Collaborative Sessions 1 and 2

Table 2 shows the percentage of green/positive, red/negative and black/free markers associated with or not associated with notes. Amongst all the markers set during both CSCL sessions ($N = 203$), only 89 were associated with notes (44%); this percentage remains relatively stable during the two sessions. The percentage of emotional markers without notes (53%) was higher than the percentage of emotional markers with notes (15%), and this difference was greater for green/positive markers ($d = 34\%$) than for red/negative markers ($d = 4\%$). Participants produced notes to accompany markers more frequently when these markers were non-emotional than when they were emotional: black/free markers were created mainly with notes (29% with and 3% without notes).

|  | Without notes | | With notes | | Total | | Change | |
|---|---|---|---|---|---|---|---|---|
|  |  |  |  |  |  |  | Without notes | With notes |
|  | No. | % | No. | % | No. | % | % | % |
| Green/positive markers | 89 | 44% | 20 | 10% | 109 | 54% | +5% | -1% |
| Red/negative markers | 19 | 9% | 10 | 5% | 29 | 14% | -6% | 1% |
| Black/free markers | 6 | 3% | 59 | 29% | 65 | 32% | +2% | -1% |
| Total | 114 | 56% | 89 | 44% | 203 | 100% | +1% | -1% |

Table 2. Number and Percentage of Red/Negative, Green/Positive and Black/Free Markers created during Synchronous Collaborative sessions Without or With Notes

*Types of Notes.* We analyzed the purpose for which students produced notes (text associated with markers) throughout the interaction with their partner. The aim of this analyze was to study the extent to which learners used notes associated to markers to self-reflect on their collaborative work when interacting (reflection-in-action processes). We classified the notes according to three categories (a note can belong to several categories): task/tool, content/note-taking and reflection (see Table 3).

Hereafter are examples of notes associated to markers, which illustrate each category:

- Tool/task-focused note: "I have difficulty hearing" (black/free marker).
- Content-focused note: "This concept comes from the developmental psychology literature" (black/free marker).
- Reflection-focused note: "The main points are clearly highlighted" (green/positive marker); "I'm feeling more anxious as time goes on" (red/negative marker); "My partner looks at her notes too much" (red/negative marker).

Table 3 shows that there were equal percentages of reflection-focused and content-focused notes. Whereas the percentage of notes used to underline relevant conceptual information remained relatively stable over the two CSCL sessions, the percentage of notes used for reflection purpose increased, while the percentage of notes used to indicate technical problems with the platform or difficulty to complete the task through the platform decreased from Session 1 to Session 2.

|  | Session 1 |  | Session 2 |  | Change |
| --- | --- | --- | --- | --- | --- |
|  | No. | % | No. | % | % |
| Tool/Task | 14 | 25% | 6 | 19% | -6% |
| Note-taking/Content | 24 | 42% | 14 | 44% | +2% |
| Reflection | 23 | 40% | 15 | 47% | +7% |

Table 3: Number and Percentage of Notes written during Synchronous Collaborative sessions according to Three Categories: Task/Tool, Content/Note-taking and Reflection

*4.3.2. Self-Reflection Reports produced after Synchronous CSCL Sessions*

We analyzed the markers used after synchronous CSCL sessions in the two reports (R1 and R2): the number, the type of markers (green/positive, red/negative, black/free markers), the authors of markers (own- and partner's markers) and the change in the use of markers in the reports between the two CSCL sessions. We also analyzed reflection sentences associated or not with markers in the two reports: their number, the reflection category to which they referred, and also their focus of reflection (self, partner or group) and their change. We also examined the extent to which the two reports differed with respect to the types of reflection sentences they contained. Finally, we studied the relationship between markers and reflection sentences associated with these markers, by analyzing which types of markers were linked to which type of sentences; and to which reflection focuses. The aim of this final analysis was to focus on the relations between reflection-in-action (putting markers during interaction) and reflection-on-action (using markers set during interaction to write the reflection reports).

*Markers used in Reflection Reports*. Table 4 shows the number of markers used in the reflection reports after both CSCL sessions. We also analyzed the type (red, green or black) and the author (self or partner) of markers. Overall, the students used 164 markers out of the 204 created during interaction. They more frequently used their own markers (68%) than their partner's markers (32%). They also used more green/positive markers (52%) than black/free markers (33%), the lowest percentage being for red/negative markers (15%).

|  | Self-markers | | Partner-markers | | Total | |
|---|---|---|---|---|---|---|
|  | No. | % | No. | % | No. | % |
| Red markers | 17 | 10% | 7 | 4% | 24 | 15% |
| Green markers | 57 | 35% | 29 | 18% | 86 | 52% |
| Black markers | 37 | 23% | 17 | 10% | 54 | 33% |
| Total | 111 | 68% | 53 | 32% | 164 | 100% |

Table 4: Number, Types (red, green or black) and Author (self or partner) of Markers used in Reflection Reports

We analyzed the change in the use of markers between report R1 (after CSCL Session 1) and report R2 (after CSCL Session 2). Table 5 shows the percentages for the different types of markers (red/negative, green/positive and black/free) used in the two reports, according to their author (self and partner). It was observed that students used fewer markers in report R2 (63) than in report R1 (101). Regarding the change in the use of self-markers between reports R1 and R2, there were (a) a decrease in the percentage of green/positive and black/free markers used (this decrease was more pronounced for black/free markers), and (b) a slight increase in the percentage of red/negative markers used. With regard to the change in the use of partner-markers between reports R1 and R2, there was (a) an increase in the percentage of green/positive and black/free markers used, and (b) a relatively small decrease in the percentage of red/negative markers used.

|  |  | R1 / Session 1 | | R2 / Session 2 | | Change |
|---|---|---|---|---|---|---|
|  |  | No. | % | No. | % | % |
| Self markers | Red markers | 10 | 10% | 7 | 11% | +1% |
|  | Green markers | 37 | 37% | 20 | 32% | -5% |
|  | Black markers | 31 | 31% | 6 | 10% | -21% |
|  | Total | 78 | 77% | 33 | 52% | -25% |
| Partner markers | Red markers | 5 | 5% | 2 | 3% | -2% |
|  | Green markers | 13 | 13% | 16 | 25% | +13% |
|  | Black markers | 5 | 5% | 12 | 19% | +14% |
|  | Total | 23 | 23% | 30 | 48% | +25% |

Table 5: Number and Percentage of Markers used in Reports R1 and R2 per Type (red, green or black) and Author (self or partner)

*Reflection Sentences of Reports*. We analyzed the number and type of reflection sentences in the two reports. Sentences that referred to the *reflection phase* were categorized as *judgment* sentences (EV: evaluation; CA: causal attributions) or *affective reaction* sentences (SA: satisfaction/affect; AD: adaptive/defensive decisions). Sentences that referred to the *forethought phase* were categorized as *task analysis* sentences (GS: goal setting; SP: strategic planning) and *motivational belief* sentences (EF: efficacy; OE: outcome expectations; IV: intrinsic value; LO: learning goal orientation).

Hereafter are examples of sentences for each category:

- Judgment sentences: "The link she makes between Vygotsky and the PZD and Piaget and his stages of development was very relevant" (EV- Partner); "I think my teammate put a positive marker at that time because she thought I answered her question in a satisfactory manner" (CA - Partner).
- Affective sentences: "Overall, pleasant exchange, good understanding, feeling, etc." (SA - Group); "I had one training session before, but the conditions were not the same" (AD - Self).
- Task analysis sentences: "I will also put more positive markers because I could have put some here" (GS – Self); "Do further research to better understand and to better explain what I have understood" (SP - Self)
- Motivational belief sentences: "I also find it hard to follow the sound and to put a marker" (EF – Self); "I think we need a few more sessions to use it at its "fair value" level" (OE – Group); "I was initially reluctant to be used as a "guinea pig" for a new program" (IV – Self); "Positive markers assist in identifying key moments to remember." (LO – Group)

As shown in Table 6, sentences in both reports mainly referred to the reflection phase

(68%), with a higher percentage for judgment sentences (45%) than for affective reaction sentences (23%). Only 18% of sentences referred to the forethought phase, with a higher proportion for task analysis sentences (12%) and motivational belief sentences (6%). 14% of all the sentences were classified as "other", and mainly described technical problems that occurred during CSCL sessions. Table 6 also depicts that there was an increase in the percentage of sentences referring to the reflection phase between reports R1 and R2. This increase mainly concerned sentences related to satisfaction (8%) and causal attribution (6%) categories. We observe a decrease in the percentage of sentences related to evaluation (-6%). The percentage of sentences referring to the forethought phase also decreased between reports R1 and R2.

|  |  | Session 1/R1 | | Session 2/R2 | | Total | | Change |
|---|---|---|---|---|---|---|---|---|
|  |  | No. | % | No. | % | No. | % |  |
| Reflection categories | EV | 58 | 29% | 25 | 23% | 83 | 27% | -6% |
|  | CA | 33 | 16% | 24 | 22% | 57 | 18% | +6% |
|  | SA | 33 | 16% | 26 | 24% | 59 | 19% | +8% |
|  | AD | 10 | 5% | 3 | 3% | 13 | 4% | -2% |
|  | Total | 134 | 66% | 78 | 72% | 212 | 68% | +6% |
| Forethought categories | GS | 5 | 2% | 3 | 3% | 8 | 3% | 0% |
|  | SP | 19 | 9% | 8 | 7% | 27 | 9% | -2% |
|  | EF | 5 | 2% | 2 | 2% | 7 | 2% | -1% |
|  | OE | 2 | 1% | 0 | 0% | 2 | 1% | -1% |
|  | IV | 5 | 2% | 2 | 2% | 7 | 2% | -1% |
|  | LO | 4 | 2% | 0 | 0% | 4 | 1% | -2% |
|  | Total | 40 | 20% | 15 | 14% | 55 | 18% | -6% |
| Other |  | 28 | 14% | 15 | 14% | 43 | 14% | 0% |
| Total |  | 202 | 100% | 108 | 100% | 310 | 100% | 0% |

Table 6: Number/Percentage of all Sentences in both Reports per Reflection/Forethought Categories

It is noteworthy that more than half of sentences in both reports (54%) were associated with a marker. We consider that a sentence is associated with a marker when it is a part of a text block that follows a marker in the reports. For instance in Figure 2, we consider that the sentences of the first two text blocks are associated with markers, while the sentences of the third text block are not.

Table 7 shows the number and percentage of sentences associated with a marker in both reports depending on reflection and forethought categories. Linked-marker sentences were mainly dedicated to reflection (89% against 1% to forethought), with a higher percentage for judgment sentences (63%) than for affective reaction sentences (26%). More precisely, linked-marker sentences referred mainly to evaluation (37%), causal attribution (26%) and satisfaction (24%) categories.

Moreover, we observed that there were more sentences linked to a marker in reports R2 than in reports R1 (+8%). Table 7 shows that the number of reflection sentences linked to a marker increased between the two reports. Compared to linked-marker sentences in reports R1, linked-marker sentences in reports R2 were more used to make causal attribution (+8%) and express satisfaction (+6%), and less used to assess the quality of collaboration (-5%). There were no linked-marker sentences that referred to forethought categories in reports R2.

|  |  | Session 1 / R1 | | | Session 2 / R2 | | | Total | | | Change |
|---|---|---|---|---|---|---|---|---|---|---|---|
|  |  | No. | % (all sentences) | % | No. | % (all sentences) | % | No. | % (all sentences) | % | % |
| Reflection categories | EV | 40 | 20% | 40% | 22 | 20% | 35% | 62 | 20% | 37% | -5% |
|  | CA | 22 | 11% | 22% | 19 | 18% | 30% | 41 | 14% | 26% | +8% |
|  | SA | 21 | 10% | 21% | 17 | 16% | 27% | 38 | 13% | 24% | +6% |
|  | AD | 3 | 1% | 3% | 1 | 1% | 2% | 4 | 1% | 2% | -1% |
|  | Total | 86 | 43% | 85% | 59 | 55% | 94% | 145 | 49% | 89% | +9% |
| Forethought categories | GS | 0 | 0% | 0% | 0 | 0% | 0% | 0 | 0% | 0% | 0% |
|  | SP | 1 | 0% | 1% | 0 | 0% | 0% | 1 | 0% | 0% | -1% |
|  | EF | 1 | 0% | 1% | 0 | 0% | 0% | 1 | 0% | 0% | -1% |
|  | OE | 0 | 0% | 0% | 0 | 0% | 0% | 0 | 0% | 0% | 0% |
|  | IV | 0 | 0% | 0% | 0 | 0% | 0% | 0 | 0% | 0% | 0% |
|  | LO | 1 | 0% | 1% | 0 | 0% | 0% | 1 | 0% | 0% | -1% |
|  | Total | 12 | 1% | 3% | 0 | 0% | 0% | 12 | 1% | 1% | -3% |
| Other |  | 3 | 6% | 12% | 4 | 4% | 6% | 7 | 5% | 9% | -6% |

| | | | | | | | | | |
|---|---|---|---|---|---|---|---|---|---|
| Total | 101 | 50% | 100% | 63 | 58% | 100% | 164 | 54% | 100% | 0% |

Table 7: Number and Percentage of Linked-Marker Sentences in both Reports per Reflection/Forethought Categories

We analyzed towards whom (themselves, their collaborative partner or their group) students' reflection processes were oriented in both reports. As it has been observed that linked-marker sentences mainly referred to reflection categories (evaluation, causal attribution, satisfaction/affect and adaptive/defensive decisions), we only consider these categories in the subsequent results. Our question also concerned whether there was a change in the focus of reflection between reports R1 and R2. In line with these questions, Table 8 displays the percentage of linked-marker sentences in reports R1 and R2 (and their change) depending on their focus (self, partner or group) and their reflection category.

As Table 8 shows, the highest percentage of linked-marker sentences in report R1 was for partner-focused sentences followed by self-focused sentences, the lowest percentage being for group-focused sentences. In report R2, the highest percentage of linked-marker sentences was for both self- and partner-focused sentences followed by group-focused sentences. More precisely, the percentage of self-focused (linked-marker) sentences increased and the percentage of group-focused (linked-marker) sentences decreased between reports R1 and R2, whereas the percentage of partner-focused (linked-marker) sentences remained stable over the two reports. There was thus a shift from group- to self-focused reflection between both reports.

Results also showed that in both reports, partner- and group-focused (linked-marker) sentences referred mainly to evaluation processes while self-focused (linked-marker) sentences referred to causal attribution and satisfaction/affect processes. There was an increase in percentage of self-focused (linked-marker) sentences for both categories (causal attribution and satisfaction/affect).

|    | Focus of sentences in R1 | | | | Focus of sentences in R2 | | | | Change of the focus | | |
|----|------|---------|-------|-------|------|---------|-------|-------|------|---------|-------|
|    | Self | Partner | Group | Total | Self | Partner | Group | Total | Self | Partner | Group |
| EV | 6%   | 21%     | 13%   | 40%   | 6%   | 19%     | 10%   | 35%   | 0%   | -2%     | -3%   |
| CA | 11%  | 8%      | 3%    | 22%   | 19%  | 11%     | 0%    | 30%   | +8%  | +3%     | -3%   |
| SA | 8%   | 10%     | 3%    | 21%   | 14%  | 10%     | 3%    | 27%   | +6%  | 0%      | 0%    |
| AD | 3%   | 0%      | 0%    | 3%    | 2%   | 0%      | 0%    | 2%    | -1%  | 0%      | 0%    |
| Total | 28% | 39%   | 19%   | 85%   | 41%  | 40%     | 13%   | 94%   | +14% | 1%      | -6%   |

Table 8: Percentage of Linked-Marker Sentences in Report R1 and Report R2 depending on their Focus (self, partner, group) and their Reflection Category (EV, CA, SA, AD)

*Markers associated to Reflection Sentences*. We analyzed the relation between markers and sentences associated with these markers. In Table 9, we focused on the author (self or partner) and type (red/negative, green/positive or black/free) of markers to which sentences were linked. We also looked at the reflection categories that were related to the self- and partner's markers.

|       | Self markers | | | | | | | | Partner markers | | | | | | | |
|-------|-----|----|-----|----|-----|----|-----|----|-----|----|-----|----|-----|----|-----|----|
|       | Red | | Green | | Black | | Total | | Red | | Green | | Black | | Total | |
|       | No. | %  | No. | %  | No. | %  | No. | %  | No. | %  | No. | %  | No. | %  | No. | %  |
| EV    | 6   | 4% | 35  | 21%| 8   | 5% | 49  | 30%| 3   | 2% | 6   | 4% | 5   | 3% | 14  | 9% |
| CA    | 7   | 4% | 5   | 3% | 13  | 8% | 25  | 15%| 2   | 1% | 4   | 2% | 9   | 5% | 15  | 9% |
| SA    | 1   | 1% | 14  | 9% | 4   | 2% | 19  | 12%| 1   | 1% | 16  | 10%| 2   | 1% | 19  | 12%|
| AD    | 0   | 0% | 1   | 1% | 1   | 1% | 2   | 1% | 1   | 1% | 1   | 1% | 0   | 0% | 2   | 1% |
| Other | 3   | 2% | 2   | 1% | 11  | 7% | 16  | 10%| 0   | 0% | 2   | 1% | 1   | 1% | 3   | 2% |
| Total | 17  | 10%| 57  | 35%| 37  | 23%| 111 | 68%| 7   | 4% | 29  | 18%| 17  | 10%| 53  | 32%|

Table 9: Number and Percentage of Linked-Marker Sentences per Type (red, green or black), Author (self or partner) and Reflection Categories (EV, CA, SA or AD).

As shown in Table 9, the highest percentages of linked-marker sentence are related to green markers (86) followed by black markers (54), the lowest being for red markers (24). More specifically, students used mainly their self-green markers followed by their self-black markers, and then their partner-green markers. The lowest percentage was for partner-red markers.

Regarding the reflection categories, evaluation (EV) mainly relied on the use of green markers, in particular on the use of self-green markers. Causal attribution (CA) sentences were mainly associated with black markers, in particular self-black markers. Satisfaction and affect (SA) were mainly expressed through sentences linked to green markers, namely with the equal use of self- and partner-green markers.

Table 10 displays the change in the use of markers between both reports depending on the author and the type of markers, as well as the reflection sub-category (EV, CA, SA or AD) to which the markers were related. The change was calculated on the difference between the percentages of markers used in report R1 and in R2.

Overall, Table 10 indicates that in reports R2, students used more the markers to make causal attribution and express satisfaction than in reports R1. The increase in causal attribution is related to an increase in the use of self-red markers and partner-black markers and a decrease in the use of self-black markers. The increase in "satisfaction/affect" category is related to an increase in partner-green markers. We observe a slight decrease in the number of sentences related to evaluation. This decrease is linked to a decrease in all types of self-markers and to an increase in partner-green and black markers.

|    | Self markers | | | | Partner markers | | | | |
|----|-----|-------|-------|-------|-----|-------|-------|-------|-------|
|    | Red | Green | Black | Total | Red | Green | Black | Total | Total |
| EV | -3% | -4%   | -3%   | -10%  | 0%  | +4%   | +3%   | +7%   | -3%   |
| CA | +6% | +3%   | -8%   | +1%   | +1% | -4%   | +9%   | +6%   | +7%   |
| SA | +2% | -4%   | -4%   | -6%   | -1% | +13%  | +1%   | +12%  | +6%   |
| AD | 0%  | +2%   | -1%   | +1%   | -1% | -1%   | 0%    | -2%   | -1%   |

Table 10: Change in the Use of Markers between Reports R1 and R2 per Author (self, partner), Type (red, green, black), and Reflection Categories (EV, CA, SA or AD)

*4.4. Discussion*

Regarding Questions 1 to 4 (markers and associated notes created during synchronous collaborative sessions), results first showed a preferential use of positive markers during interaction. This suggests that participants preferentially allocated attention toward positive emotional events of collaboration. Keeping track of positive moments in the interaction with their partner could be a self-motivation strategy used by learners to persist in performing the collaborative learning task, and also to face the negative emotions that could be experienced in response to the challenging aspects of collaboration (Järvenoja, 2010). Second (Question 2), results indicated that emotional markers – especially positive markers – were used mainly without notes. Therefore, it seems it was not necessary for learners to complement emotional markers with textual descriptions of events to which such markers referred. This may suggest that emotional markers – and in particular positive markers – convey enough information without associated text, and could serve as powerful memory cues to retrieve information about what happened during collaboration. Third (Question 3), we observed an increase between the two synchronous collaborative sessions in the percentage of positive markers created during the session. This could be explained by the fact that after the first collaborative session, learners were informed that their own markers had been made available in the retrospection room to their partner while building the self-reflection report. This information could have led learners to set more positive markers during the second collaborative session. Such a strategy may prevent their partner from experiencing non-constructive reactions in response to the visualisation of negative markers that could threaten their feeling of competence. This could also be considered a means for learners to save face during interaction (Goffman, 1955). Finally, with respect to Question 4, the notes associated to markers were used equally for note-taking and reflection-in-action purposes, with an increase for reflection in the second sessions. This result shows that – as expected – learners needed a means to support reflection during the interaction with their partner in the collaborative sessions.

Regarding Questions 5 to 9 (integration of markers and associated notes into self-reflection reports), results first showed (Question 5) that learners preferentially used their own markers – and especially their own positive markers – to self-reflect on what happened during collaboration. Therefore, learners mostly used their own perspective when making judgments about the quality of interaction with their partner, and preferentially focused on positive emotional aspects of their collaboration. Second (Question 6), we found that learners used more their partner's markers (especially positive) and less their own markers in the second reports (R2) than in the first reports (R1). This suggests that they were more likely to adopt and internalize their partner's perspective after their second experience of collaboration with their partner. When collaborative learners become more familiar with each other, one may expect that they tend to trust each other more, and feel more comfortable in their relationship. This could motivate them to take more their partner's opinions into account while self-reflecting on collaboration. Moreover, results showed a tendency for learners to focus on their partner's positive perceptions of interaction, which could be a learners' strategy to strengthen their own positive perceptions (for example, of themselves) and also to protect themselves from negative self-judgments. Third (Question 7), we were able to classify almost all the sentences using the self-regulated learning categories of Zimmerman's (2002) model. Sentences in both reports referred mainly to the *reflection* phase – with more judgment sentences than affective reaction sentences – while only a few sentences referred to the *forethought* phase. Moreover, there was an increase in reflection sentences and a decrease in forethought sentences between the two reports. This result suggests that learners' reflection processes followed a conservative rather than a progressive direction (Van der Puil et al., 2004), which lead them to identify what worked in their last interaction rather than to plan on how to improve their future group work.

With respect to Question 8, half of the sentences in both reports were associated with a marker. Linked-marker sentences mainly referred to self-reflection processes, and more precisely to evaluation processes. We observed however that compared to the first reports (R1), the second reports (R2) were used less to assess the quality of interaction, and more to make causal attributions and express satisfaction with respect to the group work. Results also showed that evaluation sentences were mainly linked to self-positive markers, and satisfaction sentences to both self- and partner's positive markers. This confirms that learners preferentially focused on positive feelings and thoughts rather than on negative ones. They also preferred giving positive evaluation based on their own perception of interaction with their partner, while they were more likely to take their partner's opinions into consideration when expressing satisfaction about the way they interacted with each other. As free markers (self and partner's markers) were principally used to make causal attribution, it appears that students used them to explain successes and failures in the collaboration process. Therefore, it seems that learners would prefer making explanatory attributions in a cognitive rather than affective mode.

With regard to the change of reflection processes, there were more self-focused sentences and less group-focused sentences in the second reports (R2) than in the first reports (R1), whereas the focus on the partner remained important and stable between the two reports. This suggests that students focused their reflection mainly on their own processes and behaviors during interaction. More precisely, it appears that the students wrote in proportion much more self-causal attribution and self-satisfaction sentences linked to markers in the second reports than in the first ones. Meanwhile, the sentences in the second reports were more linked to their partner's markers (especially more positive and free partner's markers) and less to their own markers (especially less self-free markers). This suggests that the students integrated more their partner in their own reflection-on-action, especially to justify their successes and failures and to express their satisfaction.

6. Conclusion and Future Works

In this article, we presented Visu, an innovative videoconferencing tool dedicated to both synchronous and delayed reflection in CSCL settings. As all the markers put during the synchronous collaborative sessions (one's own and those of the partner) become available in the retrospection room, Visu can provide learners with information that concern the learner, the partner and the group. To our knowledge this tool is quite unique, as there are little awareness systems that support both *reflection-in-action* and *reflection-on-action* as well as providing learners with information on their affect and motivation. That is why we adopted an exploratory approach to study the use of markers set by learning partners during interaction to reflect after the collaboration session. To sum up the main results, it appears that the students (1) used the markers equally as a note-taking and reflection means during the interaction, (2) used mainly positive markers both to reflect in and on action; (3) focused more on identifying what worked in their last interaction (conservative direction) than on reflecting on subsequent learning and task goals (progressive direction); (4) mainly used their own markers to reflect on action, with an increase in the use of their partners' markers in the second reflection reports; (5) mainly reflected on their partner in the first reflection report and more on themselves in the second report to justify themselves and to express their satisfaction.

The two last points appear very interesting and let us think that a kind of dialog is being set up between the two partners in the second reflection reports. We can suppose that a shared-regulation process did emerge in the second reflection reports, since the students seem to have self-reflected on action using the markers put in action by their learning partner. Further research is needed to test this hypothesis, and should take into account some limitations of the present study. Among these limitations is the number of CSCL sessions: a third session would have been necessary to better understand how self-reflection on collaboration evolved over time and across successive interactions with the same collaborative partner. Such a third session would also have the advantage of providing groups of students with more time for developing socially shared regulation strategies. Organizing consecutive (and also interrelated) collaboration sessions in authentic learning contexts is however a difficult task for both researchers and teachers, and such settings remain relatively rare in the CSCL field. Participants of this study were blended learning students, and the challenge was to convince them to participate in CSCL sessions from home, at night (it is usually difficult for these students to arrange a common schedule), on a topic on which to work together during consecutive CSCL sessions. Another limitation was related to the small number of distance learning students per course, a characteristic of our learning context that ensured high quality tutoring services yet also limited the number of participants in this study. Finally, the use of follow-up questionnaires and interviews would have helped to better explain the results presented in this paper. At the beginning of the face-to-face course that followed the second CSCL session, the teacher organized an informal debriefing session where students were asked about their perceptions regarding the study in which they had been involved. Students showed their interest for the Visu tool but also expressed their reluctance to work in groups (due to the constraints associated with blended learning contexts). Although interesting, the content of students' feedback was not analyzed in this study, as it did not serve its purpose. Despite these limitations, our research points to interesting directions for future research in the CSCL field, such as investigating the role of socially shared reflection and also the impact of emotional factors during the regulation of collaborative learning processes.